\documentclass[10pt,conference]{IEEEtran}

\usepackage{lipsum}
\usepackage{booktabs}
\usepackage[flushleft]{threeparttable}
\usepackage{xcolor}
\usepackage[utf8]{inputenc}
\usepackage{float}
\usepackage{amssymb}
\usepackage{amsmath}
\usepackage{ifthen}
\usepackage{multirow} 
\usepackage{caption}
\usepackage{listings}
\usepackage{svg}
\usepackage[bookmarks=false]{hyperref}
\usepackage{url,moreverb,xspace}
\usepackage[skins,breakable]{tcolorbox}

\usepackage{enumitem}
\usepackage{array,graphicx}
\usepackage{soul}
\usepackage{balance}
\usepackage{pifont}
\usepackage{etoolbox,siunitx}
\usepackage{nicefrac}
\usepackage{mathtools}

\usepackage{tikz}
\usepackage{pgfplots}
\usepgfplotslibrary{statistics}
\usetikzlibrary{fadings}
\usetikzlibrary{arrows.meta,shapes.geometric,shadows}

\usepackage{xcolor,pifont}
\newcommand*\colourcheck[1]{%
  \expandafter\newcommand\csname #1check\endcsname{\textcolor{#1}{\ding{52}}}%
}
\newcommand*\colourcross[1]{%
  \expandafter\newcommand\csname #1cross\endcsname{\textcolor{#1}{\ding{56}}}%
}
\usepackage[vlined, boxruled, linesnumbered] {algorithm2e}
\SetKw{KwBy}{by}
\SetKw{KwBreak}{break}
\SetKw{KwReturn}{return}
\def\HiLi{\leavevmode\rlap{\hbox to \hsize{\color{gray!35}\leaders\hrule height .8\baselineskip depth .5ex\hfill}}}

\definecolor{gold}{rgb}{0.83, 0.68, 0.21}

\usepackage{listings}

\lstdefinestyle{javastyle}{
  language=Java,
  basicstyle=\ttfamily\scriptsize,
  keywordstyle=\color{blue},
  commentstyle=\color{gray},
  stringstyle=\color{gold},
  numberstyle=\tiny\color{gray},
  stepnumber=1,
  numbers=left,
  numbersep=10pt,
  numbers=none, %
  tabsize=2,
  showspaces=false,
  showstringspaces=false,
  breaklines=true,
  breakatwhitespace=true,
  captionpos=b,
  morekeywords={@Override, @Test} %
}

\lstdefinestyle{normaltextstyle}{
  basicstyle=\small\ttfamily, %
  keywordstyle=\bfseries, %
  commentstyle=\itshape, %
  stringstyle=\ttfamily, %
  showstringspaces=false, %
  numbers=none, %
  frame=none, %
  framesep=0pt, %
  rulecolor=\color{gray}, %
  breaklines=true, %
  breakatwhitespace=true, %
  tabsize=2, %
  captionpos=b %
}

\newcommand*\rot{\rotatebox{90}}
\usepackage{graphicx}
\usepackage{subcaption}
\usepackage{seqsplit}

\newcommand{\gptold}{\texttt{gpt-3.5-turbo}\xspace}
\newcommand{\gptnew}{\texttt{gpt-4}\xspace}
\newcommand{\gemini}{\texttt{gemini-1.5-pro}\xspace}
\newcommand{\haiku}{\texttt{claude-3-haiku}\xspace}
\newcommand{\sonnet}{\texttt{claude-3-sonnet}\xspace}
\newcommand{\llamasmall}{\texttt{llama-3-8b}\xspace}
\newcommand{\llamabig}{\texttt{llama-3-70b}\xspace}
\newcommand{\mistralsmall}{\texttt{mistral-7b}\xspace}
\newcommand{\mistralbig}{\texttt{mistral-8x7b}\xspace}

\SetKwComment{Comment}{$\triangleright$\ }{}
\SetCommentSty{itshape}
\newboolean{showcomments}
\setboolean{showcomments}{true}
\ifthenelse{\boolean{showcomments}}
{\newcommand{\nb}[2] {
  \fcolorbox{black}{gray!20}{\bfseries\sffamily\scriptsize#1:}
  {\sf\small$\blacktriangleright$\textit{#2}$\blacktriangleleft$}
}
}
{\newcommand{\nb}[2]{}
}

\usepackage[normalem]{ulem}
\useunder{\uline}{\ul}{}

\newcommand{\head}[1]{\noindent\textbf{#1.}}

\newcounter{fcounter}
\newcommand{\curl}[1]{\footnote{\url{#1}}}

\makeatletter
\newcommand{\thickhline}{%
    \noalign {\ifnum 0=`}\fi \hrule height 1pt
    \futurelet \reserved@a \@xhline
}
\makeatother

\begin{document}
\pagestyle{plain}

\title{Improving the Readability of Automatically Generated Tests using Large Language Models}

\author{
	\IEEEauthorblockN{Matteo Biagiola \IEEEauthorrefmark{1}, 
    Gianluca Ghislotti \IEEEauthorrefmark{2},
    Paolo Tonella \IEEEauthorrefmark{1}
    }
	\IEEEauthorblockA{\IEEEauthorrefmark{1} Software Institute - Universit\`a della Svizzera italiana, Lugano, Switzerland\\
		\{matteo.biagiola, paolo.tonella\}@usi.ch}
    \IEEEauthorblockA{\IEEEauthorrefmark{2} Work done while at Universit\`a della Svizzera italiana, Lugano, Switzerland\\
		ghislottigianluca8@gmail.com}
}

\IEEEoverridecommandlockouts
\IEEEpubid{
	\parbox{\columnwidth}{XXX-X-XXXX-XXXX-X/25/\$31.00~\copyright2025 IEEE\hfill \\
		DOI XX.XXXX/ICSTXXXXX.2025.XXXXX \\
		2025 IEEE Conference on Software Testing, Verification and Validation (ICST) 
		\hfill}
	\hspace{\columnsep}\makebox[\columnwidth]{\hfill}
}

\maketitle

\IEEEpubidadjcol

\begin{abstract}
Search-based test generators are effective at producing unit tests with high coverage. However, such automatically generated tests have no meaningful test and variable names, making them hard to understand and interpret by developers. On the other hand, large language models (LLMs) can generate highly readable test cases, but they are not able to match the effectiveness of search-based generators, in terms of achieved code coverage.

In this paper, we propose to combine the effectiveness of search-based generators with the readability of LLM generated tests. Our approach focuses on improving test and variable names produced by search-based tools, while keeping their semantics (i.e., their coverage) unchanged.

Our evaluation on nine industrial and open source LLMs show that our readability improvement transformations are overall semantically-preserving and stable across multiple repetitions. Moreover, a human study with ten professional developers, show that our LLM-improved tests are as readable as developer-written tests, regardless of the LLM employed.

\end{abstract}

\begin{IEEEkeywords}
	Large Language Models,
    Software Testing,
    Readability.
\end{IEEEkeywords}

\section{Introduction} \label{section:introduction}

Automated test generation techniques have been studied in depth in  the last decades. Such techniques are very appealing as they promise to automate the test input creation process, easing the burden on software developers.

One of the prominent ways the test input creation process can be automated is by modelling the  problem via the search-based framework~\cite{harman2006search}. This methodology employs search and optimization algorithms to automatically generate test cases that maximize a certain objective, e.g., code coverage and/or bug detection, providing an efficient and scalable solution~\cite{tonella2004evolutionary}. Unit test generation for Java programs is one of the most active research areas in this field, with state-of-the-art tools such as \texttt{Evosuite}~\cite{fraser2011evosuite}, whose core generation technique has been improved over the years~\cite{fraser2013whole,arcuri2017many,panichella2018automated}. Moreover, search-based techniques have been applied to other programming languages such as Python~\cite{lukasczyk2022pynguin}, 
and Javascript~\cite{olsthoorn2024syntest}, as well as for system level testing of RESTful APIs~\cite{arcuri2018evomaster}, Web and Android applications~\cite{biagiola2017search,mao2016sapienz}. Despite their effectiveness, search-based generators have been criticized as they generate tests with low readability, making them hard to interpret (e.g., when checking the oracle, diagnosing the failures or  understanding/documenting their behaviour), and maintain~\cite{daka2015modeling,daka2017generating}.

Recently, large language models (LLMs) have been proposed as a way to address the software test creation problem. The work of Tufano et al.~\cite{tufano2020unit} pioneered this research subfield, by formalizing the test generation problem as a sequence-to-sequence translation problem. In a nutshell, \texttt{AthenaTest} exploits a Transformer architecture~\cite{vaswani2017attention} to train a language model that would translate a method under test (i.e., a focal method) into its corresponding test case. More recent approaches use pre-trained LLMs. They supply LLMs a carefully crafted prompt, which may include input/output examples (using LLMs in \textit{few-shot} mode), to generate test cases given the unit under test as context. Examples of test generators belonging to this category include \texttt{TestPilot}~\cite{schafer2023empirical}, a few-shot LLM-based test generator for Javascript, \texttt{ChatUnitTest}~\cite{chen2024chatunittest} and \texttt{ChatTester}~\cite{yuan2024evaluating}, proposing an LLM-based generation-validation-repair framework for generating Java unit tests, and \texttt{TestSpark}~\cite{sapozhnikov2024testspark}, a plugin for the \texttt{IntelliJ} IDE that supports LLM-based test generation. Multiple empirical studies in the literature show that tests generated by LLMs are more readable than tests generated by \texttt{Evosuite}~\cite{tufano2020unit,tang2024chatgpt}, as developers tend to prefer them to tests generated via coverage-guided methods. On the other hand, \texttt{Evosuite}'s tests are more effective than LLM generated tests, in terms of coverage and bug detection~\cite{siddiqa2023empirical,xie2023chatunittest,tang2024chatgpt}.

In summary, the literature suggests that coverage-guided test generators such as \texttt{Evosuite} generate effective test cases, while LLM-based generators generate more readable but less effective tests. The objective of our work is to combine the best of both worlds, i.e., the readability of LLM-based test suites while maintaining the performance of test suites generated through search-based methods. In particular, we propose to use LLMs to improve the readability of \texttt{Evosuite} tests. While existing LLM-based approaches aim to refactor the entire structure of test cases to improve their readability~\cite{gay2017generating,alshahwan2024automated}, we focus instead on identifiers and test names, as previous studies in the literature suggest that these have a large influence on the test's readability~\cite{daka2015modeling,daka2017generating}. Moreover, existing approaches do not explicitly optimize the input prompt to the LLM. Indeed, LLMs have a fixed context window, that long classes/tests may quickly saturate. Large classes/test suites, would also result in long prompts that have been shown to be detrimental for LLMs as they give rise to \textit{lost-in-the-middle} effects~\cite{liu2024lost}, where LLMs lose their capability to meaningfully attend relevant information in the middle of the prompt. We, instead, design a multi-step prompt, where we first feed the LLM the focal information of the class under test (i.e., class name, constructors, attributes and method signatures), which is kept in memory by the LLM and prefixed to each subsequent request. We then provide the LLM with each individual test to be improved, together with the method bodies of the class under test it exercises. Such individual prompts are submitted independently from each other, to keep the overall prompt short and the context window limited.

We evaluated our approach  using nine industrial and open-source LLMs. Our results show that most LLMs are able to preserve the semantics of the tests (i.e., their coverage) while improving their readability. Moreover, the readability improvements of the considered LLMs are quite stable across repetitions, despite the potential non-determinism of their output. We also conducted a human study with ten professional developers who evaluated the readability of LLM-improved tests w.r.t. developer-written tests. Results show that LLM-improved tests are equally readable as developer-written tests, regardless of the LLM used. The reliability of our stable and semantically-preserving readability transformations, as well as the comparable readability w.r.t. developer-written tests, make our approach a viable tool to be used in practice to improve the readability of coverage-guided automatically generated tests.

\section{Related Work} \label{section:related}

\subsection{Automated Test Generation} \label{section:related:automated-test-generation}

\head{Search-based Techniques} The topic of Search-based software testing has a rich literature, where numerous surveys and empirical studies have been conducted~\cite{panichella2018test,mcminn2004search,ali2010systematic}. The state-of-the-art tool for the generation of Java unit test cases is the search-based test generator \texttt{Evosuite}~\cite{fraser2011evosuite}, although \texttt{Randoop}~\cite{pacheco2007randoop}, a \textit{feedback-directed} random test generator, is often used in the literature especially as a strong baseline in Java tool competitions~\cite{jahangirova2023sbft}. Beyond Java unit test generation, search-based techniques are also used by researchers to generate test cases for RESTful APIs~\cite{arcuri2019restful} (with the tool \texttt{Evomaster}~\cite{arcuri2018evomaster}), Android GUI test cases~\cite{mao2016sapienz} (with the tool \texttt{Sapienz}) and unit tests for Python~\cite{lukasczy2023empirical} (with the tool \texttt{Pynguin}~\cite{lukasczyk2022pynguin}). Search-based techniques are very effective at generating high-coverage tests, also given their ability to focus the search on multiple areas of the search space. Indeed, one of the most recent innovation is to target all the coverage goals at once, either by aggregating the fitness of each coverage goal in a single fitness function~\cite{fraser2013whole} or by formulating the search problem as a many-objective optimization problem~\cite{panichella2018automated,arcuri2017many}, which seems to be more effective (indeed, \texttt{DynaMOSA}~\cite{panichella2018automated} is the default search algorithm in \texttt{Evosuite}).

\head{LLMs-based Test Generation} The use of LLMs in software testing is relatively new. However, from  January 2019 to October 2023, there have been more than 100 publications related to the use of LLM in software testing (more than 80\% in 2023) in both software engineering and AI venues~\cite{wang2024software}. The survey by Wang et al.~\cite{wang2024software} summarizes the state of the art on this topic, discussing the use of LLMs for software testing activities, such as unit and system test generation, oracle generation, debugging, and program repair.

Tufano et al.~\cite{tufano2020unit} describe an approach implemented in the tool \texttt{AthenaTest}, which generates test cases by solving a \textit{sequence-to-sequence} \textit{translation} problem (the tool \texttt{A3Test} by Alagarsamy et al.~\cite{alagarsamy2024a3test} is also an instance of this test generation class). The authors use an encoder-decoder transformer model (i.e., \texttt{BART}), pre-train it on a large corpus of English and Java source code, and finetune it using a translation task, where the source language is a \textit{focal} method (i.e., the method under test), and the target  is the unit test written by a human developer for that method.

More recent approaches simplify the test generation problem, by employing LLMs with \textit{zero-shots} or \textit{few-shots} prompts. 
Schafer et al.~\cite{schafer2023empirical} propose \texttt{TestPilot}, a few-shot LLM-based test generator for the API of a given Javascript project.
Plein et al.~\cite{plein2024automatic} use bug reports as prompt to a language model (in particular \texttt{ChatGPT} and \texttt{codeGPT}) to generate executable test cases.
Siddiqa et al.~\cite{siddiqa2023empirical} conduct an empirical study on the effectiveness of three language models, namely \texttt{gpt3.5-turbo}, \texttt{Codex}, and \texttt{StarCoder} for generating unit tests for Java.
Chen at al.~\cite{chen2024chatunittest,xie2023chatunittest} introduce a framework named \texttt{ChatUnitTest} for generating Java unit tests. The framework consists of a generation-validation-repair mechanism to fix errors in the generated unit test; the validation check runs a Java parser, compiles the code and runs it.
Tang et al.~\cite{tang2024chatgpt} statistically compare \texttt{ChatGPT} with \texttt{Evosuite} w.r.t. statement coverage on the \texttt{SF110} dataset, and w.r.t. bug detection on \texttt{Defects4J} projects. Results show that \texttt{Evosuite} achieves significantly higher statement coverage than \texttt{ChatGPT} (i.e., 77\% vs 55\% on average), and exposes more bugs (i.e., 55 bugs vs 44 on average). In terms of readability, which the authors measure quantitatively using code style standards as well as \textit{cyclomatic complexity}~\cite{mccabe1976complexity} and \textit{cognitive complexity}~\cite{campbell2018cognitive}, \texttt{ChatGPT} test cases do not seem to adhere to a specific code style, while they feature a low complexity that make them easy to follow. Similarly, Yuan et al.~\cite{yuan2024evaluating} propose \texttt{ChatTester}, a ChatGPT-based generator that features an intention prompt, to understand the focal method (i.e., the method under test), and a generation prompt, to generate a test for such method. 

\head{Hybrid Techniques} Researchers have also tried to combine LLMs with traditional software testing techniques to generate more effective test cases. For instance, Lemieux et al.~\cite{lemieux2023codamosa} use LLMs within the evolutionary loop of \texttt{DynaMOSA} to generate the right data to get the search unstuck. Similarly, Arghavan et al.~\cite{arghavan2024effective} use mutation testing to improve the effectiveness of test cases generated by LLMs. Such approaches do not specifically focus on readability but rather on either supporting the search-based generators in specific situations~\cite{lemieux2023codamosa}, or guiding the generation process of LLMs~\cite{arghavan2024effective}.

\subsection{Readability} \label{section:related:readability} 

\head{Code Readability}
Buse et al.~\cite{buse2009learning} construct an automated readability metric by building a classifier that predicts human readability by using a simple set of local code features.
Similarly, Campbell et al.~\cite{campbell2018cognitive} propose a cognitive complexity metric, designed to address the shortcomings of cyclomatic complexity, such as the nesting problem.
Munoz et al.~\cite{munoz2020empirical} systematically evaluated whether this metric actually captures source code understandability. Results show that cognitive complexity is a promising metric to automatically assess different aspects of code understandability, although code understandability can be measured in different ways (e.g., whether the comprehension task is completed successfully~\cite{woodfield1981effect}, time to locate and fix a bug~\cite{hofmeister2019shorter}, or perceived understandability~\cite{sedano2016code}), and it is not yet clear how those ways are related.

\head{Test Readability Improvement} Regarding software tests, researchers have focused specifically on improving the readability of automatically generated tests. Daka et al.~\cite{daka2015modeling} propose a readability metric for Java unit tests. The authors train a linear regressor on human annotated data to predict readability scores from a set of features.
Panichella et al.~\cite{panichella2016impact} propose \texttt{TestDescriber}, an approach that automatically generates a summary of the source code exercised by a certain test case.
Daka et al.~\cite{daka2017generating} focus on an automated approach to generate descriptive test names, based on the functionalities they cover. 
Roy et al.~\cite{roy2020deeptc} propose \texttt{DeepTC-Enhancer} to improve the readability of automatically generated tests. The approach works in two steps: first it generates method-level summaries for given test cases; then, it resorts to code summarization by training a deep learning model to carry out the tasks of test name and variable name predictions. The objective of the second step is to rename all the identifiers and the test name.
Delgado et al.~\cite{delgado2023interevo} adopt a different approach to test readability. In particular, they integrate readability assessments within the evolutionary loop of \texttt{Evosuite}, with the purpose of generating test cases that are more readable according to the tester's preferences. The tester gives a readability score to each test case, which the evolutionary loop takes into account to decide which tests to keep for the next generations (together with the coverage goals).

The works described above mostly adopt machine/deep learning techniques or custom heuristics to address the readability problem in automatically generated tests. Such models need to be trained on datasets of human preferences. Given the cost of labeling, the size of such datasets is limited, potentially failing to capture all the factors that might affect the subjective readability evaluations.
We instead resort to LLMs that are trained on massive corpora of source and test code, to tackle the readability task with the knowledge of many more developers. Moreover, while custom heuristics focus only on specific aspects of readability (e.g., the test names), LLMs can target the readability problem more holistically. 
The works of Gay et al.~\cite{gay2023improving} and Alshahwan et al.~\cite{alshahwan2024automated} are more related to our approach, as they both use LLMs to improve test cases. The focus of Gay et al.~\cite{gay2023improving} is to improve Python unit tests automatically generated by \texttt{Pynguin}, while Alshahwan et al.~\cite{alshahwan2024automated} propose and evaluate a tool named \texttt{TestGen-LLM} to improve  Kotlin test cases manually written for Meta's products. Despite the common objective, our work differs in several fundamental ways: (1)~our prompt is designed not to change the semantics of the test, contrary to \texttt{TestGen-LLM}~\cite{alshahwan2024automated} that specifically aims to increase the coverage of the existing test cases, and Gay et al.~\cite{gay2023improving}, whose approach potentially introduces semantic changes; (2)~we deal with the limited context window size of an LLM by designing a multi-step prompt; 
(3)~we take into account the randomness of the LLM (even at temperature zero~\cite{ouyang2023llm}) on the readability transformations, by analyzing how much the transformations vary across runs; (4)~we evaluate the performance of nine industrial and open-source LLMs, while Gay et al.~\cite{gay2023improving} only consider \texttt{gpt-4} and Alshahwan et al.~\cite{alshahwan2024automated} evaluate two LLMs internally-developed  at Meta.

\section{Motivating Example} \label{section:background}

\autoref{listing:background:stack-class} shows the \texttt{Stack} class, that we use as a motivating example. A stack has two basic functionalities, i.e., \texttt{push} and \texttt{pop}. In this example, a stack has a capacity of 3, and has no resize functionality, i.e., once the stack is full it is not possible to add new elements.

\begin{lstlisting}[style=javastyle, label={listing:background:stack-class}, caption={Class implementing the stack data structure.}]
public class Stack<T> {
    private int capacity = 3;
    private int pointer  = 0;
    private T[] objects = (T[]) new Object[capacity];
    public void push(T o) {
        if(pointer >= capacity)
            throw new RuntimeException("Stack exceeded capacity!");
        objects[pointer++] = o;
    }
    public T pop() {
        if(pointer <= 0)
            throw new EmptyStackException();
        return objects[--pointer];
    }
}
\end{lstlisting}

The \texttt{Stack} class has five branches: a branch in class constructor (not shown), two branches in the \texttt{push} method, corresponding to the condition \texttt{pointer} \verb+>=+ \texttt{capacity} being \texttt{true} or \texttt{false}, and two branches in the \texttt{pop} method for the condition \texttt{pointer} \verb+<=+ \texttt{0}. A unit test generator for this class might generate a test suite similar to the one below:

\begin{lstlisting}[style=javastyle, label={listing:background:stack-test-suite-es}, caption={Test suite for the \texttt{Stack} class generated by \texttt{Evosuite} with the branch coverage criterion. The test suite covers all  five branches of the class under test.}]
public class Stack_ESTest {
  @Test
  public void test0()  throws Throwable  {
      Stack<Integer> stack0 = new Stack<Integer>();
      try { 
        stack0.pop();
        fail("Expecting exception: EmptyStackException");
      } catch(EmptyStackException e) {
         verifyException("tutorial.Stack", e);
      }
  }
  @Test
  public void test1()  throws Throwable  {
      Stack<Integer> stack0 = new Stack<Integer>();
      Integer integer0 = new Integer(0);
      stack0.push(integer0);
      stack0.push(integer0);
      stack0.push(integer0);
      try { 
        stack0.push(integer0);
        fail("Expecting exception: RuntimeException");  
      } catch(RuntimeException e) {
         verifyException("tutorial.Stack", e);
      }
  }
  @Test
  public void test2()  throws Throwable  {
      Stack<Object> stack0 = new Stack<Object>();
      Object object0 = new Object();
      stack0.push(object0);
      Object object1 = stack0.pop();
      assertSame(object1, object0);
  }
}
\end{lstlisting}

In this example, we used \texttt{Evosuite} to generate the test suite in \autoref{listing:background:stack-test-suite-es}, a state-of-the-art search-based unit test generator for Java. In particular, we used the branch coverage criterion, such that the test generator has the objective to cover all the branches of the \texttt{Stack} class. While all branches are covered,  tests are hard to read, especially because  test names do not convey any information on what the underlying test does. Our approach aims to improve the readability of such automatically generated tests, while keeping their coverage effectiveness intact.

\section{Approach} \label{section:approach}

\begin{figure*}[ht]
    \centering

    \includegraphics[trim=0cm 12cm 2cm 0cm, clip=true, scale=0.20]
    {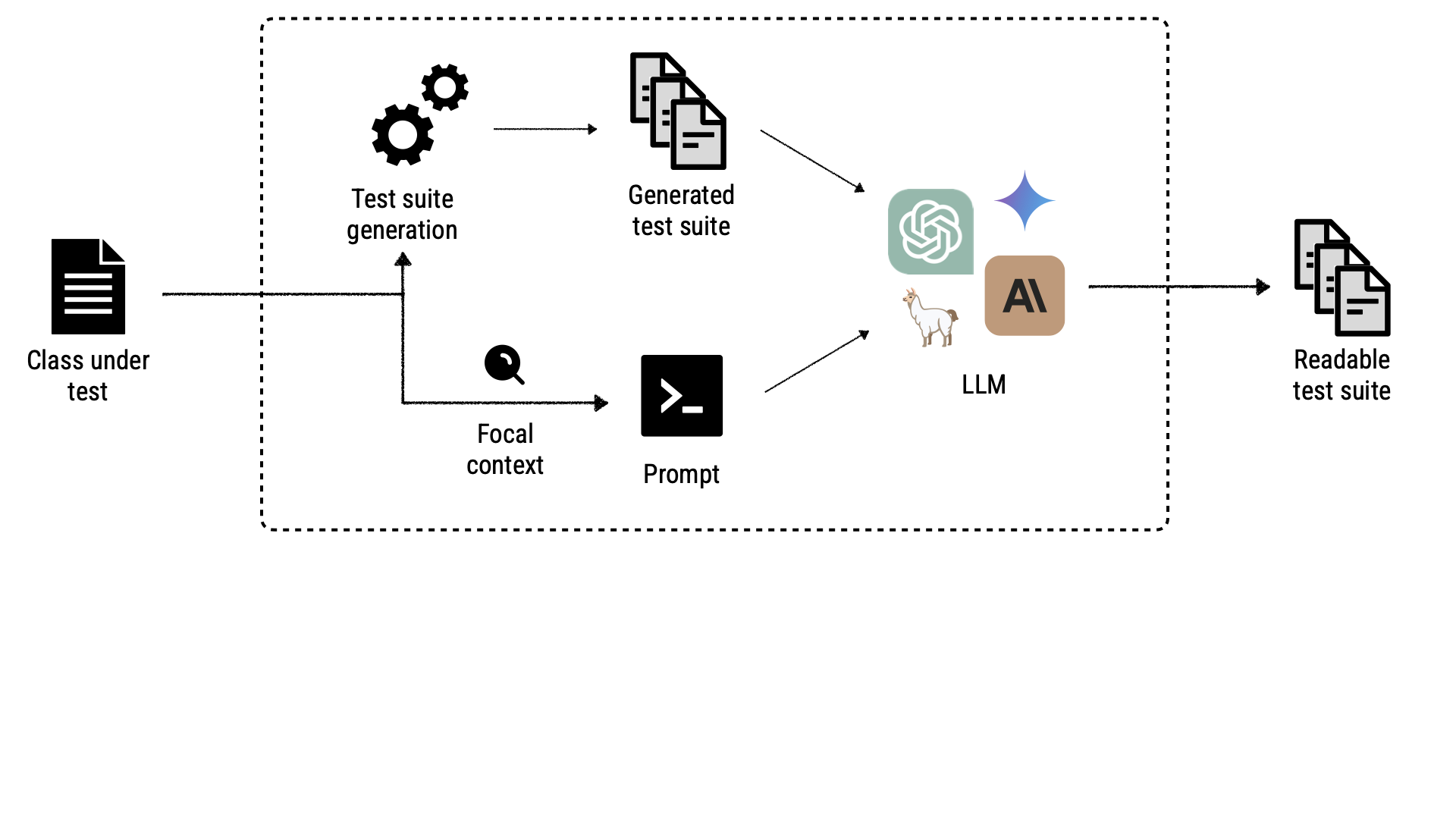}
    
    \caption{Overview of our approach to readability improvement. A test generator produces the test suite given the class under test taken as input. The \textit{focal context} of the class under test, as well as the generated test cases, make up the prompt of the LLM, improving the readability of the generated test suite.} 
    \label{fig:approach:overview} 
\end{figure*}

\autoref{fig:approach:overview} shows an overview of our approach for test case readability improvement. Our approach takes as input the class under test; then employs a test generator to automatically generate a test suite. From the test suite it extracts the individual test cases, together with the corresponding \textit{focal context} of the class under test, i.e., the functionalities of the class exercised by each test case. These two types of information make up the prompt of the LLM, which modifies each single test case and improves the readability of the generated test suite.

\subsection{Prompt Design} \label{section:approach:prompt-design}

In designing our prompts, we follow the \texttt{OpenAI} prompt engineering guidelines~\footnote{\href{https://platform.openai.com/docs/guides/prompt-engineering}{Prompt engineering guidelines (Accessed August 2024)}}, as well as established results coming from the literature~\cite{sahoo2024systematic}.
Our \textit{prompt engineering} strategy consists of splitting the readability task into multiple subtasks\footnote{\href{https://platform.openai.com/docs/guides/prompt-engineering/split-complex-tasks-into-simpler-subtasks}{Split tasks into subtasks (Accessed August 2024)}}, where we first provide the context the LLM needs to improve readability, i.e., the class under test and the goal we want to achieve, and then we feed  each single test case of the test suite that needs to be transformed. The reason is twofold: (1)~providing all available information in a single prompt, i.e., class under test and the whole test suite, and asking the LLM to improve the readability of the test suite, would be an overly complex task that could hinder the quality of the output. Moreover, long prompts can be affected by \textit{lost-in-the-middle} effects~\cite{liu2024lost}, where the model's performance degrades when it needs to access information in the middle of long prompts; (2)~LLMs have a limited context window size, measured in the number of tokens making up the input prompt. If the class under test is long, and/or the test suite contains several test cases (or few long test cases), then the input would not fit the context window size of the model, and some relevant information would be discarded. In summary, by designing a multi-step prompt, we both reduce the size of each prompt, in terms of number of tokens, and we focus the attention of the LLM on smaller tasks, avoiding long prompts.

The first prompt that we send to the LLM is the \textit{Informational prompt} (shown below). The goal of this initial prompt is to clearly state the objective of the readability task, i.e., improving the readability of test cases by only modifying the identifiers and the test names. We also specify that the readability task will be broken down into multiple steps, one per test case. To keep the size of the prompt manageable, we only include the essential information of the class under test in the initial prompt, i.e., the class name, the constructors (including their bodies), the attributes (both private and public), and the signatures of the methods (both private and public). We call such essential information \textit{focal context}, inserted into the prompt by replacing the content inside braces ``\textit{\{sc\}}''. We configured the LLM to \textit{memorize} this first prompt, such that it is prefixed to the prompts of each subsequent request.

\begin{tcolorbox}[boxrule=0.5pt, arc=2pt, toprule=0pt,  bottomrule=0.5pt, left=0mm, right=0mm, top=0mm, bottom=0mm, title=\textbf{Informational prompt}, colback=white, colframe=gray, halign=flush left, breakable]
  \textit{You are a professional Java programmer.} \\

  \textit{Your ultimate goal is to improve the readability of the test cases I will send you, particularly by modifying ONLY the identifiers, and the test name, NOT THE METHODS CALLED INSIDE THE TESTS, STATIC METHODS, OR STATIC CLASSES.} \\

  \textit{Thinking in steps:} \\

  \textit{1. Initially (this prompt), I will send you the original source code of the class under test, to give you the context and the aim of the class.} \\

  \textit{2. In the next prompts, I will send you each single test of the test suite you need to improve the readability of, as well as the source code of the original class methods that are called in the test.} \\

  \textit{Focal information of the class under test:} \\

  \textit{\{sc\}}
\end{tcolorbox}

The prompt starts with asking the model to adopt a persona\footnote{\href{(https://platform.openai.com/docs/guides/prompt-engineering/tactic-ask-the-model-to-adopt-a-persona}{Ask model to adopt persona (Accessed August 2024)}}, in this case a \textit{professional Java programmer}. The second strategy is the use of  words in capital letters to emphasize that only the identifiers and test names should be modified; indeed, we noticed from preliminary experiments that without such emphasis the LLM tends to modify method names as well. Moreover, we use the natural language statement ``Thinking in steps'' to explicitly encourage the model to follow our multi-step task. This prompt strategy is known as \textit{Chain-of-Thought} (i.e., CoT for short), and it is the foundation of all strategies related to reasoning and logic tasks. Indeed, Wei at al.~\cite{wei2022chain} propose such technique to induce a step-by-step reasoning process in the LLM, by manually detailing the logical steps such that the LLM can replicate them while solving the task. Kojima et al.~\cite{kojima2022large} found that using the phrasing ``Let's think step by step'' in the prompt would induce a similar behavior, without the cost of manually detailing the reasoning process. Finally, we use delimiters (colon and following new line) to separate the source code from the rest of the prompt\footnote{\href{https://platform.openai.com/docs/guides/prompt-engineering/tactic-use-delimiters-to-clearly-indicate-distinct-parts-of-the-input}{Use delimiters (Accessed August 2024)}}.

\begin{tcolorbox}[boxrule=0.5pt, arc=2pt, toprule=0pt,  bottomrule=0.5pt, left=0mm, right=0mm, top=0mm, bottom=0mm, title=\textbf{Improvement prompt}, colback=white, colframe=gray, halign=flush left]
  \textit{Improve the readability of the test below by modifying ONLY the identifiers, test name and variable names, NOT THE METHODS CALLED INSIDE THE TESTS, STATIC METHOD OR CALLED STATIC CLASS. The changes must not affect the functioning of the test in any way.}
  
  \textit{-----------------------------------------------------------------------------------}
  \textit{Test to modify:} \\

  \textit{\{single\_test\}}
  
  \textit{-----------------------------------------------------------------------------------}

  \textit{Knowing the source code of all the methods used in the test:} \\

  \textit{\{sc\_test\_calls\}} \\

  \textit{Answer with code only.} 
\end{tcolorbox}

The box above shows the second prompt, which we call \textit{Improvement prompt}. We send such prompt to the LLM for each test case in the generated test suite. By design,  improvement prompts are \textit{disjoint}, i.e., each improvement prompt starts a new session, discarding all the previous improvement interactions. In this way, we control the size of the prompt, which includes in each session  the informational prompt and the improvement prompt. Hence, the LLM focuses on improving the readability of one test at a time. 
The improvement prompt has three main sections: the instruction section, which instantiates the instructions in the informational prompt in the context of the current test, the source code of the current test, and the source code of the class under test methods exercised by the test.
Although our prompt design makes it difficult to reach the limit of the context window for the current LLMs, long classes and/or long test methods might still trigger a context overflow. In such situations, we proceed by first, discarding the source code of the methods exercised by the test in the improvement prompt; then, we remove the informational prompt form the memory, focusing only on improving the current test without context. In the rare cases where the improvement prompt with just the inclusion of the test to modify does not fit the prompt size, we give up improving the test and report it as-is in the final test suite. 

\begin{tcolorbox}[boxrule=0.5pt, arc=2pt, toprule=0pt,  bottomrule=0.5pt, left=0mm, right=0mm, top=0mm, bottom=0mm, title=\textbf{Remove duplicates prompt}, colback=white, colframe=gray, halign=flush left]
  \textit{These tests have the same names:} \\

  \textit{\{tests\}}
  \textit{-----------------------------------------------------------------------------------}
  \textit{Change them such that they differ, while their objectives remain clear. The content of the tests must remain exactly identical.} \\

  \textit{Answer with code only.}
\end{tcolorbox}

Finally, we check that all the tests in the final test suite have unique names. Indeed, since each improvement prompt is separated from the others, it may happen that the LLM assigns the same name to tests that are structurally or semantically similar. The \textit{Remove Duplicates Prompt} (shown above), is designed to disambiguate test names in case of duplicates. We simply provide the LLM the source code of the test cases that have the same names, and ask it to change them while keeping their content unchanged. We submit the remove duplicates prompt for each set of tests that have the same name, and proceed to query the LLM until all ambiguities are fixed (we keep track of the number of times we query the LLM, and fall back to the original test names after 3 queries). Finally, we output the improved test suite at the end of the process.

\subsection{Motivating Example}

Let us apply the proposed approach to improve the readability of the \texttt{Stack} test suite generated by \texttt{Evosuite} (\autoref{listing:background:stack-test-suite-es}). The first step, is to extract the focal context from the \texttt{Stack} class to build the informational prompt. In this case, we just copy the class in the prompt, without the bodies of the methods \texttt{push} and \texttt{pop}, replacing the ``\textit{\{sc\}}'' tag in the informational prompt. Next, for each test in the test suite, we build an improvement prompt. For instance, to improve \texttt{test1}, we replace the ``\textit{\{single\_test\}}'' tag in the improvement prompt with the source code of \texttt{test1}, and the ``\textit{\{sc\_test\_calls\}}'' tag with the source code of the \texttt{push} method of the \texttt{Stack} class.

The resulting test suite after running our approach is shown in \autoref{listing:approach:results} (we used one of the LLMs  considered in the evaluation). We observe that  test names are now meaningful and reflect what the test actually does, making the test suite more understandable and maintainable. For instance, \texttt{test1} is called \texttt{testPushCapacityExceeded} in the improved test suite, as the test indeed exercises a stack overflow. We also notice that the LLM did not change the semantics of the test cases, as it safely modified the test names and the identifiers within each test, without altering any execution flow or introducing errors.

\begin{lstlisting}[style=javastyle, label={listing:approach:results}, caption={Test suite for the \texttt{Stack} class genererated by \texttt{Evosuite} and improved by our approach.}]
@Test
public void testPopOnEmptyStack() throws Throwable {
    Stack<Integer> integerStack = new Stack<Integer>();
    try { 
        integerStack.pop();
        fail("Expecting exception: EmptyStackException");
      } catch(EmptyStackException e) {
         // no message in exception (getMessage() returned null)
         verifyException("tutorial.Stack", e);
      }
}
@Test
public void testPushCapacityExceeded() throws Throwable {
    Stack<Integer> integerStack = new Stack<Integer>();
    Integer inputInteger = new Integer(0);
    integerStack.push(inputInteger);
    integerStack.push(inputInteger);
    integerStack.push(inputInteger);
    try { 
        integerStack.push(inputInteger);
        fail("Expecting exception: RuntimeException");  
      } catch(RuntimeException e) {
         // Stack exceeded capacity!
         verifyException("tutorial.Stack", e);
      }
}
@Test
public void testPopReturnsPushedObject() throws Throwable {
    Stack<Object> objectStack = new Stack<Object>();
    Object pushedObject = new Object();
    objectStack.push(pushedObject);
    Object poppedObject = objectStack.pop();
    assertSame(poppedObject, pushedObject);
}
\end{lstlisting}

\section{Empirical Evaluation} \label{section:evaluation}

To assess the practical benefits of our readability improvement approach, we formulate the following research questions: \\

\noindent
\head{RQ\textsubscript{1} (semantic preservation)}
\textit{To what extent is the semantics of a test case preserved after readability improvements?}

\noindent
\head{RQ\textsubscript{2} (stability)}
\textit{To what extent are readability improvements stable across multiple repetitions?}

\noindent
\head{RQ\textsubscript{3} (human study)}
\textit{How do developers judge the readability of LLM-improved test cases w.r.t. developer-written test cases?}

RQ\textsubscript{1} and RQ\textsubscript{2} aim to assess whether LLMs are \textit{reliable} in the readability improvement task, making them a useful tool that can be used in practice. In particular, RQ\textsubscript{1} analyzes the extent to which the LLM changes the semantics of a test case across multiple repetitions. Although our prompt is designed not to change the semantics of a test case, the LLM might ignore the instructions and change, for instance, method calls within the test, leading to compilation errors or to exercise different execution paths of the class under test w.r.t. the original test. We promoted to the next research questions only the LLMs that never change the semantics of a test case across multiple repetitions.
In RQ\textsubscript{2} we study the \textit{stability} of the readability improvements across multiple repetitions. In other words, we analyze how much the readability improvements vary when the LLM is prompted to change the same test case multiple times. 
In RQ\textsubscript{3}, we ask developers to compare the readability of automatically generated test cases improved by LLMs, with that of developer-written test cases. Our objective with this human study is to show that LLM-improved test cases get close to the readability of developer-written test cases. Since readability is hard to measure automatically, and eventually test cases are used and interpreted by developers, we designed a human study to capture the perceived readability.

\subsection{Classes Selection} \label{section:evaluation:classes-selection}

The first step of our empirical evaluation is the selection of the classes under test. We selected classes from five well known Java projects, i.e., \textit{Apache Commons Lang} (Lang henceforth)\footnote{\href{https://commons.apache.org/proper/commons-lang}{https://commons.apache.org/proper/commons-lang (August 2024)}}, \textit{JFreeChart} (Chart henceforth)\footnote{\href{https://jfree.org/jfreechart}{https://jfree.org/jfreechart} (August 2024)}, \textit{Apache Commons Cli } (Cli henceforth)\footnote{\href{https://commons.apache.org/proper/commons-cli}{https://commons.apache.org/proper/commons-cli (August 2024)}}, \textit{Apache Commons Csv} (Csv henceforth)\footnote{\href{https://commons.apache.org/proper/commons-csv}{https://commons.apache.org/proper/commons-csv (August 2024)}}, and Google \textit{Gson} (Gson henceforth)\footnote{\href{https://github.com/google/gson}{https://github.com/google/gson (August 2024)}}. We rely on these projects because they are equipped with high quality manual test suites, which we use in the human study. 

\begin{table}[ht]
	\centering
	
	\caption{Classes selection. The table shows the number of classes remaining after each step of the selection process, except the first column (``Original'') that shows the number of classes in each project.}
	\label{table:evaluation:classes-selection}
    
    \setlength{\tabcolsep}{10pt}
    \renewcommand{\arraystretch}{1.2}
    
    \begin{tabular}{r@{\hskip 1em}r@{\hskip 0.5em}r@{\hskip 0.5em}r@{\hskip 0.5em}r}
        \toprule

        \multicolumn{1}{l}{} & \multicolumn{4}{c}{\# Classes} \\
        
        \cmidrule(r){2-5}
        
        \multicolumn{1}{l}{} & Original & Filtering & Core Selection & Readability Study \\

        \midrule

        Cli & 25 & 12 & 5 & 2 \\
        Csv & 11 & 4 & 4 & 2 \\
        Lang & 246 & 67 & 6 & 2 \\
        Gson & 73 & 23 & 6 & 2 \\
        Chart & 657 & 301 & 6 & 2 \\
        
        \bottomrule
        
    \end{tabular}

\end{table}

Since our ultimate goal is to evaluate the readability of test cases through a human study, we selected two classes per project, for a total of ten classes, in order to keep the study's size manageable. To select the classes, we first extracted from each project, and for each class within each project, the lines of code ($\textit{LOC}$), the average method length ($\textit{AML}$), in terms of number of statements, and the number of internal imports (\textit{II}), i.e., the number of classes within the respective project that are used by the specific class as dependencies. We then kept classes satisfying the following criteria: (1) not a class with a private/protected constructor and not an abstract class, (2) $\textit{LOC} \in [50, \inf]$, (3) $\textit{AML} \in [3, 20]$, (4) $\textit{II} \in [0, 10]$. The first criterion ensures that a test generator (e.g., \texttt{Evosuite}) can generate tests for those classes. The second filters out trivial classes, while the third takes care of discarding classes with no relevant functionalities (e.g., a data class with only getters and setters would have $\textit{AML} = 1$). We set an upper bound of 20 statements for the $\textit{AML}$ to exclude classes with long methods, which are likely more difficult to test and inspect than classes with shorter methods. The fourth filter ensures that the resulting classes are relatively self-contained, and do not heavily rely on other internal classes, making them easier to test and inspect. Since the number of classes was still high for some projects after applying the filters (see Column~2 of \autoref{table:evaluation:classes-selection}), we ranked the remaining classes in ascending order based on the number of internal imports (we used \textit{LOC} as a secondary sorting criterion). The first two authors then independently labeled each class as ``core''/``non-core'', to make sure that the selected classes are classes implementing core functionalities related to the project. We then selected the first six core classes (or less if after filtering the project had less classes) according to the ranking, where there was an agreement between the authors, and we made sure that \texttt{Evosuite}, the test generator we used to automatically generate tests for the selected classes, was able to generate a test suite for such classes. The number of classes after core selection is shown in Column~3 of \autoref{table:evaluation:classes-selection}. 

At this stage, we conducted a small pilot study with three PhD students from our lab, to select the two classes out of those resulting from the core selection. The objective of this pilot study was to select the two most \textit{readable} classes from a testing perspective. In the survey used for the pilot study we defined readability as the ability of a developer, different from the one who wrote the class, to design a set of test cases that adequately cover the functionalities of the class. We recommended the three PhD students to spend a maximum of five minutes to evaluate each class, and we asked them to assign a readability score, from 1 (not at all readable) to 5 (very readable), to each class within the same project.
For each project, we then summed the scores for each class, and selected the  two classes with the highest scores. The selected classes are available in our replication package~\cite{replication-package}, as well as the printout of the pilot study.

\subsection{Models Selection} \label{section:evaluation:models-selection}

We carried out the selection of the large language models to be used for  test case readability improvement in April 2024, considering two models from the following providers: \texttt{OpenAI}, \texttt{Anthropic}, \texttt{Mistral}, and \texttt{Meta}, and one model from \texttt{Google}, since only one model was available at the time of the selection. Regarding  \texttt{Anthropic}, \texttt{Mistral}, and \texttt{Meta}, we selected the models available on Amazon Bedrock\footnote{\href{https://aws.amazon.com/bedrock/}{https://aws.amazon.com/bedrock/ (September 2024)}}, which offers API access to \texttt{Claude}, \texttt{Llama3} and \texttt{Mixtral}, respectively. Regarding \texttt{OpenAI} and \texttt{Google}, we used the APIs provided by the respective providers, to access \texttt{GPT} and \texttt{Gemini} respectively. In total we selected nine models, i.e., \texttt{gpt3.5-turbo}, and \texttt{gpt-4} from \texttt{OpenAI}, \texttt{gemini-1.5-pro} from \texttt{Google}, \texttt{llama-3-8b} and \texttt{llama-3-70b} from \texttt{Meta}, 
\texttt{claude-3-haiku} and \texttt{claude-3-sonnet} from \texttt{Anthropic}, and \texttt{mistral-7b} and \texttt{mistral-8x7b} from \texttt{Mistral}.

\subsection{RQ\textsubscript{1} (semantic preservation)} \label{section:evaluation:rq1}

\subsubsection{Metrics} \label{section:evaluation:rq1-metrics}

Our study on the preservation of the test semantics after improvement is based on the reports generated by \texttt{Jacoco}\footnote{\href{https://www.eclemma.org/jacoco/}{https://www.eclemma.org/jacoco/ (September 2024)}}, an open-source toolkit for measuring and reporting Java code coverage. 
These reports contain indicate how many statements, branches, methods, and lines, were covered when running the tests. The reports provide a comprehensive and detailed overview of code coverage, allowing for an in-depth analysis of which parts of the project have been effectively tested. We define as \textit{success rate} the number of times the readability transformations performed by a given LLM are coverage-preserving, divided by the total number of repetitions. In other words, we approximate semantic-preservation with coverage-preservation.

\subsubsection{Procedure} \label{section:evaluation:rq1-procedure}

For each project, and for each class, we ran \texttt{Evosuite} with the default parameters to generate a test suite for the related classes. Then, for each LLM, and for each generated test suite, we executed our readability improvement approach ten times, to cope with the non-determinism of LLMs~\cite{ouyang2023llm}. For the LLMs requiring a temperature parameter, we set such parameter to 0, to get more deterministic results. 
Then, we compared the \texttt{Jacoco} reports when running the improved tests and those produced when running the original \texttt{Evosuite} tests. If the reports are identical, then the readability transformation is deemed successful.

\subsection{RQ\textsubscript{2} (stability)} \label{section:evaluation:rq2}

\subsubsection{Metrics} \label{section:evaluation:rq2-metrics}

We measured stability by computing the \textit{code embeddings} of the tests after readability improvement. Code embeddings are numerical representations of programs, designed to capture both the formal semantics (e.g., syntactical structure) and the informal semantics (e.g., identifier naming)  of the input code, used in tasks like code search and code completion. We embedded a pair of improved tests, i.e., $t_i$ and $t_j$, into a vector space, and we computed the \textit{cosine similarity} between such vectors. The cosine similarity varies between $-1$ and 1, and the closer it is to 1, the more similar the two test cases are. In particular, in this context $t_i$ and $t_j$ result from two different repetitions of our readability improvement approach on the same test, as we want to measure how stable the readability improvements are.

\subsubsection{Procedure} \label{section:evaluation:rq2-procedure}

We conducted stability analysis only for those LLMs that consistently preserve the original test semantics of the automatically generated test suites. For those models, and for each generated test suite, we executed our readability improvement approach five times (with the temperature set to 0) to account for non-determinism. For each test, we computed the cosine similarity by considering all the unique permutations of the test and its five versions.
We embedded each test using OpenAI text embeddings~\footnote{\href{https://platform.openai.com/docs/guides/embeddings}{OpenAI Text Embeddings (September 2024)}}, as they are known to work well with code as input. In particular, we used the \texttt{text-embedding-3-small} model. 

\subsection{RQ\textsubscript{3} (human study)} \label{section:evaluation:rq3}

\subsubsection{Procedure and Metrics} \label{section:evaluation:rq3-metrics-procedure}

For each class selected for the readability study (see Column~4 of \autoref{table:evaluation:classes-selection}), we executed both the developer-written test suite and the test suite generated by \texttt{Evosuite}, and collected the \texttt{Jacoco} reports. We ranked the tests generated by \texttt{Evosuite} for a given class by length in ascending order (to give preference to shorter over longer tests), and, for each test, we looked for developer-written tests with the same coverage profile (i.e., statements, methods, and branches) of the class under test as the automatically generated tests under analysis. If no perfect match existed, we selected the first automatically generated test with the \textit{closest match} to a developer-written test (i.e., the one with the minimum coverage differences). We followed this procedure to ensure a comparable semantics (approximated as the coverage semantics) of the automatically generated tests and developer-written tests being considered in the human study. At the end of the selection, we have two tests per class, i.e., the  automatically generated test, and the developer-written test that best matches the coverage profile of the former. On average, the automatically generated tests have a length of 5.3 statements, which is a reasonable size for test cases that need to be manually evaluated.
We then executed our readability improvement approach five times, once for each of the five LLMs that showed semantic preservation (RQ\textsubscript{1}). Hence, in total for each class we have six tests, among which five that are automatically generated and improved, and one developer-written, giving a total of 60 tests, as we have two classes per project and five projects.

We designed the survey for the human study to be at most 30-minutes long, to avoid that fatigue would affect the readability assessment of the tests. As we have two classes per project, we split the survey in two parts, i.e., a group of developers would evaluate the tests related to the first class of each project, while the other group of developers would evaluate the tests related to the second class of each project. In this way, a single developer is asked to evaluate 30 tests in 30 minutes, i.e., one test per minute, which was deemed feasible in preliminary trials.

We organized each part of the survey into five blocks of questions, i.e., one per project, where in each block we placed the six tests related to the class under test for the specific project. We then asked developers to assign a score to each of the test, from $-2$ (the test code is very unreadable) to $2$ (the test code is very readable).
In each block of questions, we also added an optional text-box where developers could justify and comment their scores. To avoid a learning effect, we randomized the order in which the tests are presented in the survey (printouts of the survey are in our replication package~\cite{replication-package}).

To conduct the study, we hired ten professional developers, i.e., five for the first survey and five for the second one, on \textit{Upwork}\footnote{\href{https://www.upwork.com}{https://www.upwork.com (September 2024)}}. We selected this platform because it is a global freelancing platform that facilitates remote collaboration between clients and professionals, and because it was used in other studies in software engineering~\cite{jahangirova2021empirical}.
We posted a fixed-priced job with a payment of 20 USD
\footnote{The client assumes the costs of the platform, which consists of a flat-fee per contract initiation and VAT.}. 
We calculated the price as 30 USD per hour, based on the average salary of a software developer in the US\footnote{\href{https://www.payscale.com/research/US/Industry=Software_Development/Hourly_Rate}{Payscale SD hourly rate (September 2024)}}, plus 5 USD for completing the qualification task. We clarified explicitly in the payment terms of the job post that payment would be due only if the qualification task and the survey were successfully completed.

The  qualification task consists of finding a functional bug we seeded in a Java class  selected from the \texttt{CodeDefenders} benchmark~\cite{fraser2019gamifying}. In particular, we selected the \texttt{Lift} class among the available 12 classes, as its functionalities are relatively easy to understand in a short amount of time, which is compatible with a qualification test (the source code of the \texttt{Lift} class is in our replication package~\cite{replication-package}). In total, we interviewed 11 developers, of which one did not pass the qualification test. All the others were able to successfully write a test that exposes the functional bug.

Based on the information we collected during the survey, the ten developers we hired had different experiences in software development, ranging from 1-3 years (1 developer) to more than 5 years (7 developers). Most of them (6 developers) reported to be at an intermediate level regarding their software testing skills, with 3 being expert and one beginner.

\subsection{Results}

\subsubsection{RQ\textsubscript{1} (semantic preservation)} \label{section:evaluation:results-rq1}

\begin{table*}[ht]
	\centering
	
	\caption{Results for RQ\textsubscript{1} and RQ\textsubscript{2}. The table shows the average success rate (Columns~1--9) and the average cosine similarity (Columns~10--14) of each LLM across five repetitions, for each project, considering all the test cases and classes.}
	\label{table:evaluation:rq1-rq2}
 
    \setlength{\tabcolsep}{8.3pt}
    \renewcommand{\arraystretch}{1.2}
    
    \begin{tabular}{lrrrrrrrrrrrrrr}
        \toprule

        & \multicolumn{9}{c}{\textbf{RQ\textsubscript{1}(semantic preservation)}} & \multicolumn{5}{c}{\textbf{RQ\textsubscript{2} (stability)}} \\

        \cmidrule(r){2-10}
        \cmidrule(r){11-15}

        & \multicolumn{9}{c}{\textbf{Success Rate (\%)}} & \multicolumn{5}{c}{\textbf{Cosine Similarity}} \\

        \cmidrule(r){2-10}
        \cmidrule(r){11-15}

        & \rot{\gptold} & \rot{\gptnew} & \rot{\gemini} & \rot{\llamasmall} & \rot{\llamabig} & \rot{\haiku} & \rot{\sonnet} & \rot{\mistralsmall} & \rot{\mistralbig} & \rot{\gptold} & \rot{\gptnew} & \rot{\gemini} & \rot{\haiku} & \rot{\sonnet} \\

        \midrule

        Cli & 100.00 & 100.00 & 100.00 & 0.00 & 50.00 & 100.00 & 100.00 & 0.00 & 20.00 & 0.87 & 0.95 & 0.94 & 0.90 & 0.94 \\
        Csv & 100.00 & 100.00 & 100.00 & 20.00 & 60.00 & 100.00 & 100.00 & 0.00 & 40.00 & 0.83 & 0.82 & 0.92 & 0.88 & 0.89 \\
        Lang & 100.00 & 100.00 & 100.00 & 40.00 & 0.00 & 100.00 & 100.00 & 0.00 & 10.00 & 0.96 & 0.98 & 0.94 & 0.88 & 0.88 \\
        Gson & 100.00 & 100.00 & 100.00 & 0.00 & 80.00 & 100.00 & 100.00 & 0.00 & 30.00 & 0.97 & 0.99 & 0.88 & 0.89 & 0.94 \\
        Chart & 100.00 & 100.00 & 100.00 & 40.00 & 60.00 & 100.00 & 100.00 & 0.00 & 0.00 & 0.96 & 0.95 & 0.88 & 0.81 & 0.87 \\

        \midrule

        \multicolumn{1}{r}{\textit{Avg}} & 100.00 & 100.00 & 100.00 & 20.00 & 50.00 & 100.00 & 100.00 & 0.00 & 20.00 & 0.92 & 0.94 & 0.91 & 0.87 & 0.90 \\
        
        \bottomrule

    \end{tabular}
\end{table*}

Columns~1--9 of \autoref{table:evaluation:rq1-rq2} show the average success rates, in terms of semantic preservation, of the nine evaluated LLMs across five repetitions, per project for all the selected tests in the respective classes. 
Results show that  different models exhibit significant differences in their ability to preserve the test semantics. By looking at the \textit{Avg} row, the \gptold, \gptnew, \gemini, \haiku, and \sonnet models achieved an average success rate across all projects of 100\%, highlighting their reliability and suitability for the readability task. In contrast, the \mistralsmall, \llamasmall, \mistralbig, and \llamabig models showed significantly lower success rates, with average success rates considering all projects of 0\%, 20\%, 20\%, and 50\%, respectively, indicating a lower ability to preserve the test semantics when improving their readability. Among models that are not always semantically-preserving, bigger models tend to be better than their smaller counterparts (i.e., \mistralbig has a higher success rate than \mistralsmall, and \llamabig is better than \llamasmall). Most of the times the readability improvements of such LLMs are not semantically-preserving due to  compilation errors. By inspecting the tests improved by such LLMs, we noticed that the most frequent semantic-breaking change is the modification of the name of the class when instantiating the constructor. Another common semantic-breaking change is the name of the class methods called in the test.

\begin{tcolorbox}[boxrule=0pt,frame hidden,sharp corners,enhanced,borderline north={1pt}{0pt}{black},borderline south={1pt}{0pt}{black},boxsep=2pt,left=2pt,right=2pt,top=2.5pt,bottom=2pt]
    \textbf{RQ\textsubscript{1} (semantic preservation)}: \textit{
        Overall, most considered LLMs  (5 out of 9) are able to follow the instructions in the prompt and do not change the semantics of the tests they are improving.
    }
\end{tcolorbox}

\subsubsection{RQ\textsubscript{2} (stability)} \label{section:evaluation:results-rq2}

Columns~10--14 of \autoref{table:evaluation:rq1-rq2} show the average cosine similiarity values for each LLM across five repetitions, per project, considering  all the test cases and classes. All the models seem to produce very similar tests across repetitions, as the average cosine similarity across projects ranges from a minimum of 0.87 (\haiku), to a maximum of 0.94 (\gptnew). 

\begin{figure*}[ht]
    \centering

    \includegraphics[trim=0cm 24cm 0cm 0cm, clip=true, scale=0.26]
    {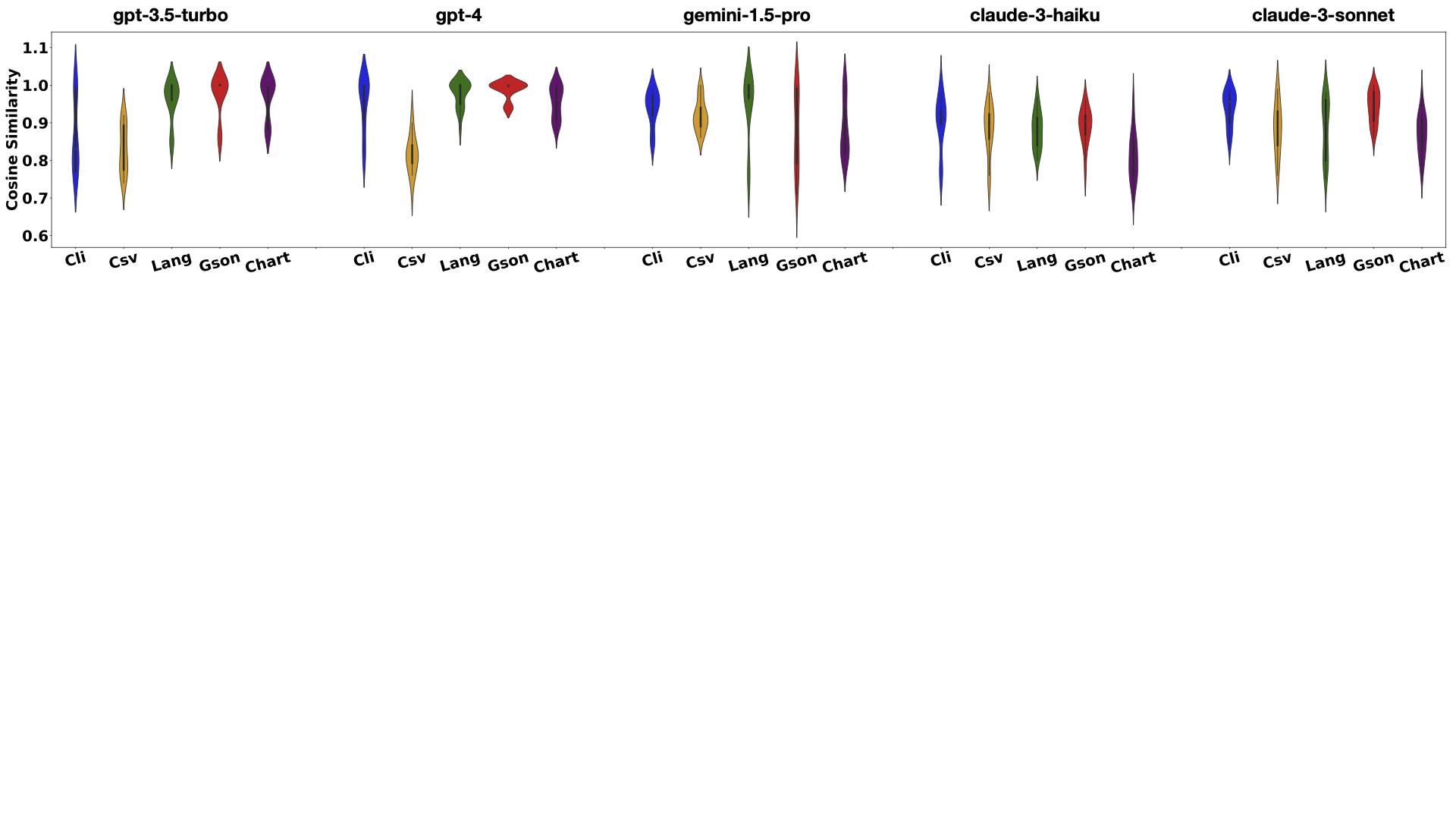}
    
    \caption{Results for RQ\textsubscript{2}. The figure shows the distributions of cosine similarity values for each LLM, for each project, considering all the test cases and classes. The x-axis shows the different projects for each LLM, while the y-axis shows the cosine similarity, which ranges from $-1$ to $1$.} 
    \label{fig:evaluation:rq2-stability} 
\end{figure*}

\autoref{fig:evaluation:rq2-stability} shows the distribution of cosine similarities for each LLM and project. The $x$-axis shows the five projects, while the $y$-axis shows the cosine similarity values. We observe that the \gptnew and \gptold LLMs show generally narrow distributions for the Lang, Gson and Chart projects. On the contrary, \gemini seems to be more stable in the remaining projects, i.e., Cli and Csv, while its results vary more on Lang, Gson and Chart. The \haiku and \sonnet LLMs show similar patterns across projects. Considering the variability per project, almost all LLMs seem to be quite stable in Cli and Gson classes (except \gptold and \gemini respectively), while Csv and Lang classes are those where LLMs exhibit higher variability.

\begin{tcolorbox}[boxrule=0pt,frame hidden,sharp corners,enhanced,borderline north={1pt}{0pt}{black},borderline south={1pt}{0pt}{black},boxsep=2pt,left=2pt,right=2pt,top=2.5pt,bottom=2pt]
    \textbf{RQ\textsubscript{2} (stability)}: \textit{
        Overall, all LLMs produce tests with improved readability that have a high similarity between each other, hence exhibiting high stability.
    }
\end{tcolorbox}

\subsubsection{RQ\textsubscript{3} (human study)} \label{section:evaluation:results-rq3}

\begin{figure}[ht]
    \centering
    
    	\includegraphics[trim=0cm 0cm 0cm 0cm, clip=true, scale=0.19]
    	{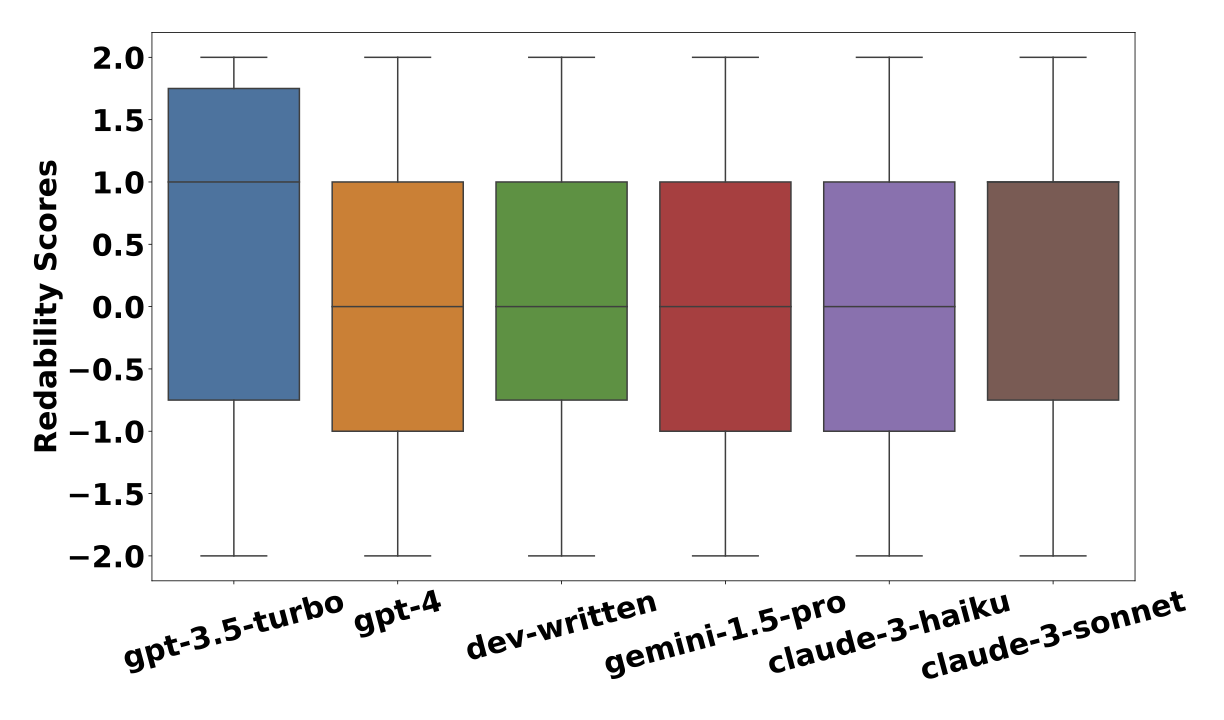}
    
    \caption{Distributions of readability scores for each LLM across projects (RQ\textsubscript{3}). The x-axis shows the different LLM models, plus the tests written by developers (\texttt{dev-written}); the y-axis shows the readability scores, ranging from $-2$ (the test code is very unreadable) to $2$ (the test code is very readable).} 
    \label{fig:evaluation:rq3-human-study} 
\end{figure}

\autoref{fig:evaluation:rq3-human-study} shows the distributions of readability scores given by developers for each LLM and across all the classes for all the projects. The $x$-axis shows the models we considered for the study, as well as the developer-written tests (\texttt{dev-written}). The $y$-axis shows the readability score ranging from $-2$ to 2. Overall, we observe that the boxplots for all the models, including the \texttt{dev-written} tests, overlap for the majority of their respective interquantile ranges. Although the medians of the \gptold and \sonnet models seem to be higher than the other models (including the \texttt{dev-written} tests), the respective differences with the \texttt{dev-written} tests are not statistically significant according to the Wilcoxon test~\cite{wilcoxon1945individual} (with a significance level at $\alpha < 0.05$). 
Due to the small sample size (50 = 10 developers $\times$ 5 projects), the statistical power $\beta$ did not reach the conventional threshold of 0.8, commonly used to accept with confidence the null hypothesis of no significant difference. 
However, the boxplots show a clear picture, with substantial readability score overlap and no indication of lower readability w.r.t. developer-written tests.

\begin{tcolorbox}[boxrule=0pt,frame hidden,sharp corners,enhanced,borderline north={1pt}{0pt}{black},borderline south={1pt}{0pt}{black},boxsep=2pt,left=2pt,right=2pt,top=2.5pt,bottom=2pt]
    \textbf{RQ\textsubscript{3} (human study)}: \textit{
        Overall, developers judge the readability of developer-written tests similar to that of LLM-improved tests.
    }
\end{tcolorbox}

\subsection{Threats to validity} \label{section:evaluation:threats}

\head{Internal validity} Internal validity threats might come from the way the empirical study was carried out. To ensure a fair comparison between LLM-improved tests and developer-written tests, we selected tests with a comparable coverage. Moreover, we selected known and documented classes with high-quality test suites, to avoid comparing LLM-improved tests with developer-written tests that have a low readability. Such classes were chosen with a systematic procedure, by first adopting automated filters and then resorting to an internal survey with PhD students, to manually assess the suitability of those classes for the readability task.
Another internal validity threat comes from the design of the prompt. We mitigated this threat by following existing guidelines for prompt engineering. Other internal threats come from the way we designed the survey for the human study. We mitigated these threats, by (1)~splitting the survey into two parts, to avoid a fatigue effect that would result from the assessment of a high number of tests; (2)~randomizing the order in which tests are presented to developers during the survey, to avoid the learning effect.

\head{External validity} External validity threats concern the generalizability of our results. We mitigated these threats by selecting five well-known Java projects, and nine LLMs from different providers.
Results could also be affected by the population of developers registered on Upwork. We mitigated this threat by designing a qualification test to assess the skills of developers before completing the survey.

\head{Conclusion validity} Threats to conclusion validity may come from random variations in the experiments. We mitigated this threat by studying semantic preservation and stability using five repetitions of our readability improvement approach, as LLMs are notoriously non-deterministic~\cite{ouyang2023llm}.

\head{Construct validity} Construct validity threats are related to the choice of inappropriate metrics. To measure test semantic preservation we used the coverage information provided by \texttt{Jacoco} during the execution. Although tracking the whole computation state would have been more precise, coverage is a computationally efficient proxy for semantics, and it is widely used in the literature~\cite{fraser2011evosuite}.

\section{Conclusion and Future Work} \label{section:conclusion}

Search-based unit test generators have shown to be very effective at exercising the functionalities of software classes. However the generated  tests have low readability, as the test names do not reflect the semantics of the execution, and identifiers are generic and hard to interpret. Large language models (LLMs), on the other hand, generate readable tests, but with an overall lower effectiveness. In this paper, we proposed to combine the effectiveness of search-based unit test generators with the ability of LLMs to improve and manipulate natural language and source code. Our approach takes as input a class under test, generates a test suite, and uses an LLM to improve each test by focusing on the test names and identifiers. Results show that LLM-improved tests maintain the semantics of the original tests, while improving their readability, making it comparable to that of developer-written tests. In our future work, we plan to extend our approach with Retrieval Augmented Generation~\cite{lewis2020retreival}, to take into account additional knowledge when the LLM is prompted for readability improvements.

\section{Data Availability} \label{section:data-availability}

Our replication package is publicly available~\cite{replication-package}, making our results reproducible.

\balance
\bibliographystyle{IEEEtran}
\bibliography{paper}

\end{document}